\documentclass{article}
\usepackage{amssymb,amsmath,color}
\usepackage{url,afterpage,epsfig,float}
\usepackage{citesort}

\newcommand{\pr}{\text{Pr}}
\newcommand{\aic}{\text{AIC}}

\newcommand{\lk}{\text{lk}}
\newcommand{\ie}{\it i.e. \rm}
\newcommand{\eg}{\it e.g. \rm }

\title{Probability Models for Degree Distributions of Protein Interaction Networks}
\author{Michael P.H. Stumpf and Piers J. Ingram\\ Centre for Bioinformatics, Department of Biological Sciences\\ Imperial College London, SW7 2AZ, London}
\begin{document}

\maketitle

\begin{abstract}
The degree distribution of many biological and technological networks has been described as a power-law distribution. While the degree distribution does not capture all aspects of a network, it has often been suggested that its functional form contains important clues as to underlying evolutionary processes that have shaped the network. Generally, the functional form for the degree distribution has been determined in an ad-hoc fashion, with clear power-law like behaviour often only extending over a limited range of connectivities. 
Here we apply formal model selection techniques to decide which probability distribution best describes the degree distributions of protein interaction networks. Contrary to previous studies this well defined approach suggests that the degree distribution of many molecular networks is often better described by distributions other than the popular power-law distribution. This, in turn, suggests that simple, if elegant, models may not necessarily help in the quantitative understanding of complex biological processes.\\
\end{abstract}

\section{Introduction}
Technological advances seen in molecular biology and genetics increasingly provide us with vast amounts of data about genomic, proteomic and meta\-bolo\-mic network structures\cite{Ito2000,Wagner2001,Qin2003}. Understanding the way in which the different constituents of such networks, --- proteins in the case of protein interaction networks (PIN) --- interact is believed to yield important insights into basic biological mechanisms\cite{Agrafioti2005,Yook2004}. For example the extent of phenotypic plasticity allowed for by a network, or levels of similarity between molecular networks in different organisms, presumably depend at least to some extent on topological (in a loose sense of the word) properties of networks. 
\par
The degree distribution, the number of nodes $n(k)$ that have $k$ connections to other nodes is one of the important characteristics of networks \cite{Albert2002,Newman2003b}. The observed degree distribution can be used to define an empirical probability distribution. If $N$ is the total number of nodes in the network then the probability that a node has $k$ edges is defined via $\pr(k)= n(k)/N$; both $\pr(k)$ and $n(k)$ are often referred to as the degree distribution. It is widely understood that the degree distribution is only one, and by no means the most important summary statistic of a network. Other frequent measures are the clustering coefficient, network diameters as well as motif frequencies and graph spectra, but so far degree distributions are generally the most widely studied characteristic. 
\par     
It has frequently been suggested that natural and technological networks, including PINs, show scale-free behaviour and that the degree distribution follows a power-law $
\pr(k) = k^{-\gamma}/\zeta(\gamma)$, 
where $\zeta(\gamma)$ is Riemann's zeta-functions which is defined for $x>1$ and diverges as $\gamma\rightarrow 1\downarrow$; for finite networks, however, it is not necessary that the value of $\gamma$ is restricted to values greater than 1. Indeed most, if not all, empirical degree distributions tail off slower than exponentially: real biological, technological and social networks tend to have a few nodes with many more connections than would be expected for classical, or Erd{\"o}s-R\'enyi, random graphs\cite{Bollobas1998}. But so far treatments have focused on fitting power-law (or heuristically derived finite-size versions of power-laws) to the observed degree distributions.
\par
The notion of ``scale-free'' has a precise mathematical meaning. If a function $f(k)$ is scale-free then the ratio $f(\alpha k)/f(k)$ depends only on $\alpha$ but not on $k$.  Most empirical degree distributions lack this property, at least globally. If plotted on a log-log plot then many degree distributions do indeed take on the shape of a straight line, at least over a range of connectivities, but never over the whole range of connections. 
\par
 Here we will determine which probability model best describes the degree distribution over the whole range of degrees for a number of different networks. Our trial distributions are chosen from among the distributions which are known to occur for theoretical networks models (from graph theory or statistical physics), supplemented by some well-known probability distributions with fat tails. In addition to the Poisson distribution ($\exp(-\lambda)\lambda^k/k!$ for all $k\ge 0$; {\bf M1}) and the power-law ($k^{-\gamma}/\zeta(\gamma)$; {\bf M4}) we will also consider the exponential ($C\exp(-k/\bar{k})$ for all $k\ge 0$ with normalizing constant $C$; {\bf M2}), the Gamma distribution ($k^{\gamma-1}e^{-k}/\Gamma(\gamma)$ for all $k\ge 0$; {\bf M3}), the log-normal ($C e^{-\ln((k-\theta)/m)^2/(2\sigma^2)}/[(k-\theta)\sigma\sqrt{2\pi}]$ for all $k\ge 0$; {\bf M5}) and the stretched exponential ($C\exp(-\alpha k/\bar{k}) k^{-\gamma}$ for $k>0$; {\bf M6}). As we will see, power-law distributions do not always perform better than the other fat-tailed distributions. 

\section{Likelihood analysis of the degree distribution of a network}
We briefly introduce the basic statistical concepts employed later. These can be found in much greater detail in most modern statistics texts such as \cite{Davison_book}. 
 The likelihood of a statistical model, $M$, given the observed data $D=\{D_1,D_2,\ldots,D_n\}$ is defined via 
\begin{equation} 
L(M) \propto \pr(D|M) = \prod_{i=1}^n \pr(D_i|M).
\label{likelihood}
\end{equation}
Taking logarithms on both sides of Eqn. (\ref{likelihood}) yields the log-likelihood, and since the proportionality constant in Eqn. (\ref{likelihood}) may depend on the data but not on the underlying statistical model (\ie the probability distribution), we  may write
\begin{equation}
\lk(M) = \sum_{i=1}^n\log(\pr(D_i)|M).
\label{loglikelihood}
\end{equation}
\par
The model $M$ is parameterized by a set of $\nu$ parameters $\mathbf{\theta}=\{\theta_j\}$ and the maximum likelihood estimates (MLE), $\hat{\theta}$, of the parameters are the values for which the expressions on the left-hand side of Eqns. (\ref{likelihood}) and (\ref{loglikelihood}) become maximal. For these values the observed data is more likely to occur than for any other parameters. 
\par

 Ultimately, however, we would like to be able to determine how much better, for example, a scale-free model is at explaining an observed degree distribution than a Poisson model. 
For non-nested models we have to employ an information criterion such as the Akaike information criterion (AIC) \cite{Akaike1983,Burnham1998} or the Bayesian information criterion (BIC). The AIC for a model $M_j$ is defined via 
\begin{equation}
\aic_j = 2(-\lk(M_j)+v_j)
\end{equation}
where $v_j$ is the number of parameters needed to define model $M_j$. 
The model with the minimum AIC is chosen as the best model and the information criteria balance a model's power against its complexity\cite{Akaike1983,Burnham1998}. In order to compare different models we define the relative differences $
\Delta^{\aic}_j = \aic_j-\min(AIC)$.
This in turn allows us to estimate the relative likelihoods of the models $
{\cal L}(M_j)\propto\exp(-\Delta^{\aic}_j/2)$.
Normalizing these relative likelihoods yields the so-called Akaike weights $w_j$
\begin{equation}  
w_j =\frac{\exp(-\Delta^{\aic}_j/2)}{\sum_{j=1}^Jexp(-\Delta^{\aic}_j/2)}.
\end{equation}
The Akaike weight $w_j$ can be interpreted as the probability that model $M_j$ (out of the $J$ alternative models) is the best model given the observed data. The relative support for one model over another is thus given by the ratio of their respective Akaike weights. The Akaike weight formalism is very flexible and has been applied in a range of contexts, including the assessment of confidence in phylogenetic inference \cite{Strimmer2002}. The equivalent quantities for the BIC, which arises as a limit of formal Bayesian model selection are called Schwarz weights and can be interpreted in the same manner. In the next section we will apply this formalism to PIN data from five species and estimate the level of support for each of the models discussed above.

\section{Results}
\begin{table}
\centering
\begin{tabular}{|c||c|c|c|c|c|c|}\hline
Model&M1&M2&M3&M4&M5&M6\\
\hline Nr. of parameters&1&1&1&1&2&3\\
\hline 
{\it D.melanogaster}&-38273&-20224& -29965&-18520&-17835&{\bf -17820}\\
{\it C.elegans}&-9017&-5975&-6071&-4267&-4328&{\bf -4248}\\
{\it S.cerevisiae}&-24978&-14042&-20342&-13459&{\bf -12713} &-12759\\
{\it H. pylori}&-2552&-1776&-2052&-1595&{\bf -1527}&-1529\\
{\it E.coli}&-834&-884&-698&-799&{\bf -659}&-701\\
\hline
\end{tabular}
\vskip1mm
\caption{Log-likelihoods for the six degree distributions discussed in the text for PIN data collected from five model organisms. The likelihoods of the models with the highest Akaike weights  (which is always $\omega_i\approx1$) are indicated in bold.} 
\end{table}
In table 1 we show the likelihoods for the degree distributions calculated from PIN data collected in five model organisms\cite{xenarios_NAR2000} (the protein interaction data was taken from the DIP data-base; http://dip.doe-mbi.ucla.edu). We find that the standard scale-free model never provides the best fit to the data; in three networks ({\it C.elegans, S.cerevisiae} and {\it E.coli}) the lognormal distribution (M5) explains the data best. In the remaining two organisms the stretched scale-free model provides the best fit to the data. The bold likelihoods correspond to the highest Akaike weights. Apart from the case of {\it H.pylori} (where $\max(w_j)=w_5\approx 0.95$ for M5 and $w_6\approx 0.05$) the value of the maximum Akaike weight is always $>0.9999$. 
\afterpage{
\begin{center}\begin{figure}
 \begin{center}
\epsfig{file=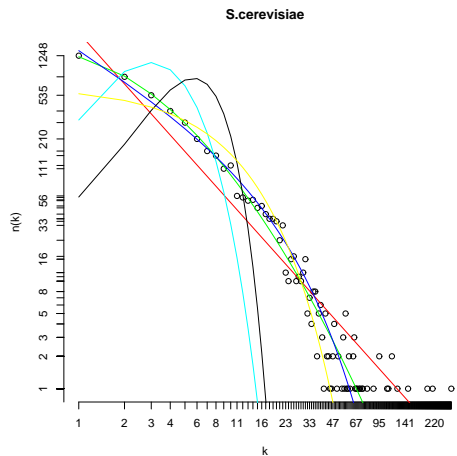,height=6.5cm}
\caption{Yeast protein interaction data (o) and best-fit probability distributions: Poisson (\textcolor{black}{\bf ---}), Exponential (\textcolor{yellow}{\bf ---}), Gamma (\textcolor{cyan}{\bf ---}), Power-law (\textcolor{red}{\bf ---}), Lognormal (\textcolor{green}{\bf ---}), Stretched exponential (\textcolor{blue}{\bf ---}). The parameters of the distributions shown in this figure are the maximum likelihood estimates based on the real observed data. }
\end{center}\end{figure}\end{center}}
\par
For the yeast PIN the best fit curves (obtained from the MLEs of the parameters of models M1-M6) are shown in figure 1, together with the real data. Visually, log-normal (green) and stretched exponential (blue) appear to describe the date almost equally well. Closer inspection, guided by the Akaike weights, however, shows that the fit of the lognormal to the data is in fact markedly better than the fit of the stretched exponential. But the failure of quickly decaying distributions such as the Poisson distribution, characteristic for classical random graphs \cite{Bollobas1998} to capture the behaviour of the PIN degree distribution is obvious. 
\par
Figure 2 shows only the three curves with the highest values of $\omega_i$, which apart from {\it E.coli} are the lognormal, stretched exponential and power-law distributions; for {\it E.coli}, however, the Gamma distribution replaces the power-law distribution. These figures show that, apart from {\it C.elegans} the shape of the whole degree distribution is not power-law like, or scale-free like,  in a strict sense. Again we find that lognormal and stretched exponential distributions are hard to distinguish based on visual assessment alone. Figures 1 and 2, together with the results of table 1, reinforce the well known point that it is hard to choose the best fitting function based on visual inspection. It is perhaps worth noting, that the PIN data is more complete for {\it S.cerevisiae} and {\it D.melanogaster} than for the other organisms.

\par
The standard scale-free model is superior to the lognormal only for {\it C.elegans}.
The order of models (measured by decreasing Akaike weights) is M6, M5, M4, M2, M3, M1 for {\it D.melanogaster}, M6, M4, M5, M2, M3, M1 for {\it C.elegans}, M5, M6, M4, M2, M3, M1 for {\it S.cerevisiae} and {\it H.pylori}, and M5, M3, M6, M4, M2, M1 for {\it E.coli}. Thus in the light of present data the PIN degree distribution of {\it E.coli} lends more support to a Gamma distribution than to a scale-free (or even stretched scale-free) model. There is of course, no mechanistic reason why the gamma distribution should be biologically plausible but this point demonstrates that present PIN data is more complicated than predicted by simple models. Therefore  statistical model selection is needed to determine the extent to which simple models really provide insights into the intricate architecture of PINs. While none of the models considered here will be anywhere close enough to the true model
it is apparent from the present study 
that there is as yet no simple probability model that could explain all PIN degree distributions satisfactorily, including the flexible stretched exponential \cite{Sornette2003}. 
\par
We have also determined the likelihoods of the data for two heuristic finite-size versions of the powerlaw\cite{Wagner2001,Dorogovtsev_book}: their performance is markedly increased compared to M4 but still behind that of the log-normal and stretched exponential (data not shown). Moreover, we simulated 1000 scale-free networks with 400 nodes (using the mathematically well defined LCD construction) and found that the AIC always favoured the scale-free model over the lognormal distribution; measured by the AIC, however, stretched exponential and powerlaw have comparable power at explaining the data with a slight advantage for the more flexible model M6 (the average Akaike weight for the lognormal model is smaller than $10^{-60}$). Thus the power of model-selection extends even to  relatively small networks and it is therefore unlikely that the above observations can be explained solely by finite size effects.
\par 
For completeness we note that model selection based on BIC results in the same ordering of models as the AIC shown here. 
\afterpage{\begin{figure}
\begin{center}
\epsfig{file=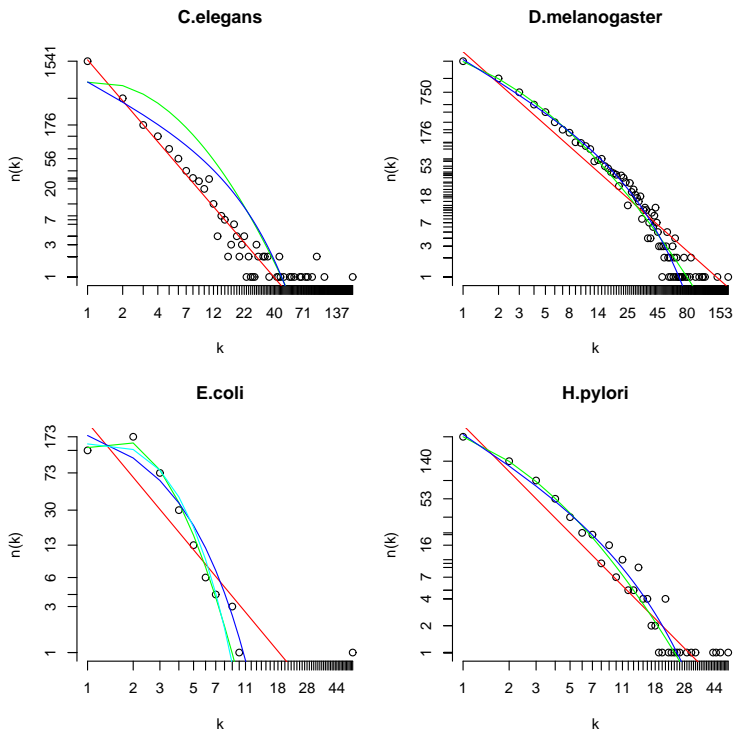,height=10cm}
\caption{Degree distributions of the protein interaction networks (o) of {\it C.elegans}, {\it D.melanogaster}, {\it E.coli} and {\it H.pylori}. The power-law (\textcolor{red}{\bf ---}), lognormal (\textcolor{green}{\bf ---}) and stretched exponential (\textcolor{blue}{\bf ---}) models are shown for all figures; for {\it E.coli} the gamma distribution (\textcolor{cyan}{\bf ---}), which performs better (measured by the Akaike weights) than either scale-free and the stretched exponential distributions.}
\end{center}
\end{figure}}\par

%
%

\section{Discussion}

\par
Several approaches have been developed which aim to describe networks more completely than is possible by the degree distribution (see \eg \cite{Milo2002,Middendorf2004,Berg2004}) and this will continue to be an important area of research for some time. The aims of the present study, however, was to determine if the conclusions which have been drawn from analysis of degree distributions are statistically sound. This is relevant as the degree distribution is still seen as the litmus test of whether a network is scale-free; there certainly is no evidence of that for PINs.
\par
There still remains the point as to how much we can expect real networks to conform to simple models. Intriguingly it has often been admitted that the degree distribution exhibits power-law behaviour only over small ranges of $k$, perhaps because of finite-size effects. Quite generally, and as pointed out by Jensen\cite{jensen_book}, powerlaw behaviour should extend over several decades in order to be meaningfully interpreted (and inferred). In our analysis of finite-size networks we found however, that the approach used here is able to "distinguish" reliably between the fat-tailed distributions and will call the scale-free model in the vast number of cases. 
\par
Here we investigated only a small number of degree distributions and we do not claim that they are anywhere similar enough to the true distribution to be meaningful representations of the underlying network. We do find, however, that fat-tailed degree distributions, like power-laws and the lognormal and stretched exponential distributions offer quantitatively better descriptions of the data than distributions which decay rapidly. However, the popular scale-free models often do a disappointing job at capturing the quantitative (to some extent even the qualitative) characteristics of protein interaction networks compared to lognormal and stretched exponential distributions.
\par
It has to be kept in mind that model selection differs considerably from hypothesis tests \cite{Burnham1998}: it determines which model, from a given set of models, can best explain the data. Formal model selection does not, however, test if data was drawn from a particular probability distribution. Crucially, if the true model is not included in the trial set $\{M_j\}$ then model selection will nevertheless determine a rank ordering of the models $M_j$ even though they may be very different from the true model. 
\par
 There are a number of reasons which may contribute to the failure of theoretical models to describe real degree distributions: (i) PIN data is notoriously noisy and plagued by false-positive and false-negative results. Curated databases like DIP, however, try to keep the number false-positive interactions low. (ii) Models are often formulated for, or solved in, the thermodynamic limit (network size $N\longrightarrow \infty$) while real networks are finite. Finite size effects can, however, often be straightforwardly incorporated into the formalism (either heuristically or through explicit numerical modelling).
(iii) The evolutionary process is much more complicated, contingent and erratic, than the models from statistical physics: yeast, for example, has undergone a whole genome duplication some $10^8$ years ago\cite{wolfe1997}. Such events no doubt affect PIN organization but in a way that is difficult to model statistically. (iv) Most networks studied to date do not represent the whole network but are in fact smaller subnets sampled from the whole network. Depending on the sampling process the subnet can differ radically from the overall network. Finally, (v) biological networks may not be best described by static graphs. No doubt interactions between proteins depend on external stimuli and are conditional on the presence or absence of other chemical entities. 
\par
Technological advances, as well as refined statistical tools, will over time address the first point. The second point is mainly technical/computational. The remaining three issues, however, are of considerable interest to theoretical biologists and statistical physicists. Statistical network ensembles  that are more flexible than the simple scale-free models \cite{Barabasi1999b} need to be investigated more systematically. But  even if a mechanistic model is not correct in detail, a corresponding statistical ensemble may nevertheless offer important insights\cite{Burda2001}, and the Akaike formalism will help to keep model complexity at an acceptable, though necessary, level (our results indicate there is no danger of over-fitting present PIN data as yet).
\par

In summary, we have shown how statistical methods can be applied to network data. These methods suggest that simple, if elegant, models from statistical physics need to be refined in order to gain quantitative insights into network evolution. 
We believe that the statistical models employed here will also be useful in helping to identify more realistic ensembles of network models. 

\section*{Acknowledgments}

We thank the Wellcome Trust for a research fellowship (MPHS) and a research studentship (PJI). We thank Bob May, Martin Howard and Carsten Wiuf for helpful discussions and comments on this manuscript.

\bibliographystyle{unsrt}
\bibliography{/home/michael/bibliography/mstbibnet}

\begin{thebibliography}{10}

\bibitem{Ito2000}
T.~Ito, K.~Tashiro, S.~Muta, R.Czawa, T.~Chiba, M.~Nishizawa, K.~Yamamoto,
  S.~Kuhara, and Y.~Sakaki.
\newblock Towards a protein-protein interaction map of the budding yeast: A
  comprehensive system to examine two-hybrid interactions in all possible
  combinations between the yeast proteins.
\newblock {\em PNAS}, 97:1143, 2000.

\bibitem{Wagner2001}
A~Wagner.
\newblock The yeast protein interaction network evolves rapidly and contains
  few redundant duplicate genes.
\newblock {\em Mol.Biol.Evol.}, 18(7):1283--1292, 2001.

\bibitem{Qin2003}
Hong Qin, Henry H~S Lu, Wei~B Wu, and Wen-Hsiung Li.
\newblock Evolution of the yeast protein interaction network.
\newblock {\em Proc Natl Acad Sci U S A}, 100(22):12820--4, Oct 2003.

\bibitem{Agrafioti2005}
I.~Agrafioti, J.~Swire, I.~Abbott, D.~Huntely, Butcher S., and M.P.H. Stumpf.
\newblock Comparative analysis of the saccaromyces cerevisiae and
  caenorhabditis elegans protein interaction networks.
\newblock {\em BMC Evolutionary Biology}, 5:23, 2005.

\bibitem{Yook2004}
Soon-Hyung Yook, Zoltán~N Oltvai, and Albert-László Barabási.
\newblock Functional and topological characterization of protein interaction
  networks.
\newblock {\em Proteomics}, 4(4):928--42, Apr 2004.

\bibitem{Albert2002}
R~Albert and AL~Barabasi.
\newblock Statistical mechanics of complex networks.
\newblock {\em Rev.Mod.Phys.}, 74(1):47--97, 2002.

\bibitem{Newman2003b}
MEJ Newman.
\newblock The structure and function of complex networks.
\newblock {\em SIAM Review}, 45(2):167--256, 2003.

\bibitem{Bollobas1998}
B.~Bollob{\'a}s.
\newblock {\em Random Graphs}.
\newblock Academic Press, 1998.

\bibitem{Akaike1983}
H.~Akaike.
\newblock Information measures and model selection.
\newblock In {\em Proceedings of the 44th Session of the International
  Statistical Institute}, pages 277--291, 1983.

\bibitem{Burnham1998}
{K.P.} Burnham and {D.R.} Anderson.
\newblock {\em Model Selection and Multimodel Inference}.
\newblock Springer, 1998.

\bibitem{Strimmer2002}
K.~Strimmer and A.~Rambaut.
\newblock Inferring confidence sets of possibly misspecified gene trees.
\newblock {\em P.Roy.Soc.Lond. B}, 269:127--142, 2002.

\bibitem{xenarios_NAR2000}
I.~Xenarios, D.~Rice, L.~Salwinski, M.~Baron, E.~Marcotte, , and D.~Eisenberg.
\newblock Dip: the database of interacting proteins.
\newblock {\em Nucl.Acid.Res.}, 28:289--291, 2000.

\bibitem{Sornette2003}
D~Sornette.
\newblock {\em Critical Phenomena in Natural Sciences}.
\newblock Springer, 2003.

\bibitem{Milo2002}
R~Milo, S~Shen-Orr, S~Itzkovitz, N~Kashtan, D~Chklovskii, and U~Alon.
\newblock Network motifs: Simple building blocks of complex networks.
\newblock {\em Science}, 298(5594):824--827, 2002.

\bibitem{Middendorf2004}
M.~Middendorf, E.~Ziv, C.~Adams, J.~Hom, R.~Koytcheff, C.~Levovitz, G.~Woods,
  L.~Chen, and C.~Wiggins.
\newblock Discriminative topological features reveal biological network
  mechanisms.
\newblock {\em BMC Bioinformatics}, 5:181, 2004.

\bibitem{Berg2004}
J~Berg and M~L\"assig.
\newblock Local graph alignment and motif search in biological networks.
\newblock {\em PNAS USA}, 101(41):14689--14694, 2004.

\bibitem{jensen_book}
Henrik~F. Jensen.
\newblock {\em Self-organized criticality}.
\newblock Cambridge University Press, 1998.

\bibitem{wolfe1997}
K.~Wolfe and D.~Shields.
\newblock Molecular evidence for an ancient duplication of entire yeast genome.
\newblock {\em Nature}, 387:708--713, 1997.

\bibitem{Barabasi1999b}
AL~Barabasi and R~Albert.
\newblock Emergence of scaling in random networks.
\newblock {\em Science}, 286(5439):509--512, 1999.

\bibitem{Burda2001}
Z.~Burda, J.~Diaz-Correia, and A.~Krzywicki.
\newblock Statistical ensemble of scale-free random graphs.
\newblock {\em Phys.Rev. E}, 64:046118, 2001.

\end{thebibliography}

\end{document}